# Direct visualization of electromagnetic wave dynamics by laser-free ultrafast electron microscopy

**Authors:** Xuewen Fu[1,2,*], Erdong Wang[3], Yubin Zhao[4], Ao Liu[4], Eric Montgomery[4], Vikrant J. Gokhale[5], Jason J. Gorman[5], Chunguang Jing[4], June W. Lau[6], Yimei Zhu[1,*]

**Affiliations:**

[1]Condensed Matter Physics and Material Science Department, Brookhaven National Laboratory, Upton, New York 11973, USA

[2]School of physics, Nankai University, Tianjin 300071, China

[3]Collider-Accelerator Department, Brookhaven National Laboratory, Upton, NY 11973, USA

[4]Euclid Techlabs, LLC, 365 Remington Blvd, Bolingbrook, USA

[5]Microsystems and Nanotechnology Division, National Institute of Standards and Technology, Gaithersburg, MD 20899, USA

[6]Materials Science and Engineering Division, National Institute of Standards and Technology, Gaithersburg, MD 20899, USA

*Corresponding author. E-mail: zhu@bnl.gov; xwfu@nankai.edu.cn

**ABSTRACT**

**Integrating femtosecond (fs) lasers to electron microscopies has enabled direct imaging of transient structures and morphologies of materials in real time and space, namely, ultrafast electron microscopy (UEM). Here we report the development of a laser-free UEM offering the same capability of real-time imaging with high spatiotemporal resolutions but without requiring expensive fs lasers and intricate instrumental modifications. We create picosecond electron pulses for probing dynamic events by chopping a continuous beam with a radiofrequency (RF)-driven pulser, where the repetition rate of the electron pulses is tunable from 100 MHz to 12 GHz. A same broadband of electromagnetic wave is enabled for sample excitation. As a first application, we studied the GHz electromagnetic wave propagation dynamics in an interdigitated comb structure which is one of the basic building blocks for RF micro-electromechanical systems. A series of pump-probe images reveals, on nanometer space and picosecond time scales, the transient oscillating electromagnetic field around the tines of the combs, and time-resolved polarization, amplitude, and nonlinear local field enhancement. The success of this study demonstrates the feasibility of the low-cost laser-free UEM in real-space visualizing of dynamics for many research fields, especially the electrodynamics in devices associated with information processing technology.**

**MAIN TEXT**

Modern electron microscopy, due to the picometer wavelength of high energy electron beam and the recent advances in aberration-correction and direct detector technologies, enables imaging of matter with atomic resolution (*1-3*). Together with the progress made in electron crystallography, tomography, cryo-single-particle imaging, and other analytical methods (*4-9*), it has become a central tool in many fields of research from materials science to biology (*10-12*). In a typical conventional electron microscope, the electron beam is produced by a thermionic or a field emission process without any control over its temporal behavior. With such an electron source images are either static or taken at long time intervals due to the limitation of the millisecond refresh rate of the conventional electron detectors. To investigate the reaction paths in physical and chemical transitions beyond the detector limitations, a high temporal resolution is required for the advanced electron microscope.

Controlled release of a pulsed electron beam is a proven method to produce time-resolved electron microscopy for studying the elementary dynamical processes of structural and morphological changes, i.e., ultrafast electron microscopy (UEM) (*13, 14*). Several methods have been developed to achieve a pulsed electron beam in an electron microscope, such as electrostatic beam blanker (*15-17*) and laser-actuated photoemission (*18-23*), which makes nanosecond (ns) and sub-picosecond (ps) (respectively) dynamics accessible. For the former, the intrinsic ns duration of the electron pulse largely restricts the temporal resolution. For the latter, further optimization of the photoemission using microwave compression (*24, 25*), terahertz compression (*26-30*) or photon-gating (*31, 32*) can extend the temporal resolution into the deep femtosecond (fs) regime, which has found vast applications in studying the transient structures and morphologies of inorganic and organic materials (*13, 18, 33*). Therefore, the laser-actuated photoemission scheme is currently the primary method for UEM. However, there are several barriers for achieving laser-actuated photoemission: fs lasers can be bulky and expensive, instrumental modifications are intricate, and beam fluctuation is an intrinsic problem due to the inevitable laser pointing instabilities. Moreover, the excitation for samples generally is limited to the same fs laser source, which not only produces significant heating, but also has been largely prohibited to the study of device physics, especially the high-frequency electrodynamics. Electrodynamics of devices, particularly in the GHz range, is fundamentally important because the standards of global data transfer (WiFi, 4G, 5G and processor clock speeds etc.) and radiofrequency (RF) micro-electromechanical systems (MEMS) are almost all in the GHz range (*34, 35*).

It has been proposed that chopping a continuous electron beam through the combination of a resonant RF deflection cavity and a small aperture is a promising alternative to create short electron pulses for the implementation of a laser-free UEM (*36-39*), where the continuous beam is periodically swept across the aperture, resulting in a pulsed beam conserving the original peak brightness and energy spread. The advantage is that no intrusive alteration to the electron source and fs lasers are required. However, the resonant RF deflection cavity can only operate at a particular high resonance frequency that is sensitive to the ambient thermal fluctuation, and requires a very high RF power for actuation. Furthermore, the resonant RF deflection cavity will induce a dual pulsed-beam due to the inevitable creation of two pulses with different divergence angles in each RF period. So far, no ultrafast pump-probe imaging or diffraction by the proposed RF cavity-driven UEM has been achieved.

Here, we report the development of a laser-free UEM by integrating a homemade RF-driven electron beam pulser to create short electron pulses with a tunable repetition rate from 100 MHz up to 12 GHz, which provides the capability to record ultrafast image and diffraction of structure transitions. With optimization of the input RF power and frequency for the pulser, a ~10 ps time resolution is achieved in our instrument. Moreover, the same broadband tunable RF signal can be used to provide sample excitation. As a first demonstration of its capability for studying ultrafast dynamics, we carried out a pump-probe study on electromagnetic (EM) wave propagation dynamics in a microstrip specimen consisting of two interdigitated combs, which is one of the basic building blocks for RF MEMS (*40*). Under a 5.25 GHz EM wave excitation, the stroboscopic imaging reveals, in real time and space, unambiguous temporal oscillating EM fields around the tines of the combs with time-dependent polarization direction and strength. Moreover, a clear nonlinear local field enhancement is observed at the corners of each tine. Combined numerical simulations and experimental results we uncovered the electrodynamics of a GHz EM wave propagation in the microstrip specimen, which is fundamentally essential to the operation of most information processing devices and currently inaccessible by other imaging methods.

The conceptual design of the laser-free UEM is schematically presented in Fig. 1A, which outlines the interfacing of the RF-driven pulser system with a transmission electron microscope (TEM). Fig. 1B shows a picture of our prototype laser-free UEM system based on a 200 keV TEM (JEOL JEM-2100F, a Lorentz TEM with a Schottky field emission source) (*41*). The pulser, inserted between the electron gun and the microscope's first condenser lens, consists of two traveling-wave metallic comb stripline elements with a small chopping aperture between them (insets of Fig. 1A and 1B). The details of the design have been described elsewhere (*39, 42, 43*). Briefly, the top stripline element is an electron beam modulator (K1) while the bottom one is a demodulator (K2), and both operate in the traveling wave mode. The input RF signals to K1 and K2 have the same frequency and are phase locked to a master oscillator with their amplitude (i.e. power) and their relative phase are digitally tunable. When the pulser is activated by a RF signal with the frequency of $f_0$, a sinusoidal EM field is generated in the modulator K1, introducing an oscillating transverse momentum kick (in the *x-y* plane, where z is the optic axis) to the coming continuous electron beam. The beam begins to oscillate in the *x-y* plane and sweep across the chopping aperture. The chopping aperture partitions the continuous beam into periodic electron pulses with a repetition rate of $2f_0$, because two pulses are created in each RF period. Further downstream, as the pulses enter the demodulator K2, a phase and amplitude optimized oscillating EM field established in K2 fully compensates the transverse momentum induced by K1 to reduce the emittance growth and energy spread of the pulses, preserving the spatial and temporal coherence. Note that, the K2 compensation plays a critical role for resolving the dual pulsed-beam issue due to the modulator induced transverse momentum on the chopped pulses (*36, 37*), which is crucial to realize the ultrafast pump-probe measurements. Since both K1 and K2 operate in the traveling wave mode, a broadband EM field with a frequency ranging from 50 MHz to 6 GHz can be established in our current design. Thus due to the frequency doubling, the repetition rate of the electron pulse is tunable from 100 MHz to 12 GHz. Unless otherwise specified, the RF frequency of $f_0 = 2.625$ GHz is used for all the experimental data presented in this work.

To perform a true ultrafast pump-probe experiments, the sample should be excited at the same repetition rate as the probe electron pulses. For the EM wave excitation configuration (Fig. 1A), since the repetition rate of the pulsed beam is $2f_0$, we split a small part of RF signal (~10% of the power) from the RF source of K1 by a coupler and doubled its frequency to $2f_0$ with a frequency doubler. After passing through a downstream amplifier, a phase shifter is used to control the time delay (i.e. phase delay) between the excitation EM wave and the probe electron pulse. Finally, we use a specially designed TEM sample holder with broad bandpass and low power loss to efficiently deliver the EM wave to the sample (Fig. 1A). Moreover, using advanced laser-RF synchronization technologies with little pulse jitter (*44, 45*), the excitation for samples is extensible to laser pulses, namely, the laser-triggered pump-probe experiments can be performed as well with our instrument.

The rationale behind the design of the RF-driven pulser is to realize laser-free UEM preserving the original modalities of the TEM when the RF activation is off. To test the performance of the TEM after integrating the pulser, we recorded a set of imaging and diffraction results under the same condition at both continuous beam (conventional TEM) mode and pulsed beam mode (Fig. 2). At the maximum magnification (200 kX) of this Lorentz TEM with a field-free weakly excited objective lens, the bright field images of gold nanoparticles at both modes are comparable in the intensity profile and contrast (Fig. 2A and 2E). For the out-of-focus Fresnel phase imaging, both modes show the similar phase contrast on the magnetic vortex in a circular ferromagnetic Permalloy disk (Fig. 2D and 2H). For the diffraction tests, diffraction patterns of gold nanoparticles (Fig. 2B and 2F) and a $VO_2$ single crystal (Fig. 2C and 2G) were recorded in both modes, which exhibit no obvious change other than the expected intensity decline in the pulsed beam mode. The comparable quality of imaging and diffraction between the pulsed beam mode and the continuous beam mode illustrates the good performance and versatility of our prototype laser-free UEM.

The temporal resolution of the laser-free UEM is mainly determined by the duration of the chopped electron pulses, which depends on the duty cycle of the chopped electron beam and can be altered independently by changing the input RF power ($P_{rf}$) and/or the chopping aperture size. Theoretically, the chopped pulse duration is given by $\tau = \gamma m_e(d + r)/4qE_0 l$ (*36*), where $\gamma$ is the Lorentz factor, $m_e$ the electron mass, *d* the diameter of the chopping aperture, *r* the diameter of the electron beam at the position of the chopping aperture, *q* the elementary charge, $E_0$ the EM field in K1, and *l* the distance between K1 and the chopping aperture. It can be retrieved by measuring the decrease in total electron counts per second with the full beam illuminating on the camera when the beam waist at the chopping aperture is smaller than the aperture diameter (*37*). As presented in Fig. 3A, the measured electron pulse duration decreases with increasing the voltage amplitude ($U_0$) of the input RF source for the modulator K1 and follows the theoretically expected behavior $\tau \propto 1/U_0 \propto 1/\sqrt{P_{rf}}$ (fit in Fig. 3A) (*36*). At the maximum input RF power of ~8 W and using the minimum chopping aperture of ~25 $\mu$m in our current design, a shortest pulse duration of ~10 ps is achieved. In principle, using higher input RF power and/or a smaller chopping aperture could achieve shorter and even sub-ps or fs electron pulses (*37*), which is promising to further improve the temporal resolution.

To demonstrate the ultrafast pump-probe measurement capability of our laser-free UEM, we carried out ultrafast imaging study on the EM wave propagation dynamics in a microstrip consisting of two interdigitated combs (Fig. 3C and Fig. S1). Understanding electrodynamics in microstrips is important as the oscillating currents and fields are fundamental to the operation of almost any information processing devices (*27*). However, direct visualizing the electrodynamics at GHz frequencies in microstrips has not been achieved so far to the best of our knowledge due to the lack of proper transient imaging technology. The sample was fabricated on a silicon on insulator (SOI) wafer and using a typical SOI microfabrication process (Materials and Methods), and was designed to match the wave impedance at around 5 GHz, which is the frequency regime for the advanced 5G wireless communication technologies. Specifically, the total length of each comb is 1.25 mm, and the tine pitch on both combs is 20 $\mu$m. The width and length of each tine is $w = 4.5$ $\mu$m and $L=75$ $\mu$m, respectively, with a gap between the interleaved tines of $g = 3$ $\mu$m (Fig. S1). The thickness of the tines along the beam-path direction is $D_z = 25$ $\mu$m. In the experiment, the input terminal of one comb was excited by a 5.25 GHz EM wave with a power of ~1.0 W (Fig. 3A), while the output end of the same comb was terminated with a 50 $\Omega$ load to eliminate signal reflections (Fig. 3C). The other comb was held at ground potential. The wavelength of the GHz wave for excitation is about 11.5 cm in vacuum, but only about 3 cm in our microstrip due to the large relative permittivity (~12) of the silicon layer. Thus the 1.25 mm comb sample spans less than 5% of a full wave. Under the GHz wave excitation, the intentional local EM fields around the tines of the interdigitated combs would give a deflection to the imaging electron pulse in *x-y* plane and result in a change in the image. Since the electron pulse duration is nearly 19 times shorter than the cycle (~190 ps) of the excitation EM wave, it allows to take images at a series of specific delay times for time-frozen electrodynamics in the sample.

First time-resolved images of EM propagation in the interdigitated comb structure acquired at a magnification of 1200 X are shown in Fig. 3D (Movie S1), where a set of typical snapshots (two ground tines and one active tine in between them) at different delay times obtained from the area indicated by the blue dashed box in Fig. 3C are presented, revealing a pronounced temporal oscillation or breathing of the tines in the time-frozen images. With the delay time increasing from zero ps (time zero was set at a delay point when the beam has no deflection), the width of the middle active tine gradually shrinks first and then broadens, while the width of the two ground tines gradually broadens first and then shrinks in alternation. More specifically, the retrieved width variation (along *x* direction) versus time of the two ground tines follows a sinusoidal function (red dots in Fig. 3E, only shows the data for one of the ground tines), while that of the active tine follows a cosine function (blue dots in Fig. 3E). Through the fitting it is found that the width variations of both the active and ground tines show a similar amplitude (~90 nm) and have an identical frequency of 5.25 ± 0.02 GHz (Fig. 3E), which is consistent with the frequency of the GHz wave for excitation.

For better analyzing the experimental result, we denote the spatiotemporal electric and magnetic fields around the tines as $E(x, y, z, t)$ and $B(x, y, z, t)$, the electron pulse velocity as $v_e$, and the frequency of the excitation wave as $f$. Considering the following conditions: (1) $D_z/v_e \ll 1/f$, the electron penetration time through the sample is much shorter than the cycle of the excitation wave, where $D_z$ is the thickness of the sample along the beam-path direction; (2) the pulse duration is nearly 19 times shorter than the periodicity of the excitation wave; (3) the effects of magnetic fields are negligible compared to that of the electric fields for the specimen geometry (*27*); and (4) the pulsed beam is collimated at the sample. The approximate change in beam divergence angles $\alpha_{x,y}$ after penetrating the sample at each position in the beam and at a delay time of $t$ is given by $\alpha_{x,y}(x,y,t) \approx qE_{x,y}(x,y,t)D_z/m_e v_e^2$ (*27*). At a specific delay time $t$, if the electric field vectors (in the *x-y* plane) around a tine point outwards from the tine's surface, each ray in the pulsed beam subjects to an field-dependent momentum kick towards the tine's surface and thus a change of divergence angle $\alpha_{x,y}(x,y,t)$, resulting in a beam deflection towards the tine's surface and a shrinking of the tine in the image; in contrast, if the electric field vectors point towards the tine's surface, both the momentum kick and the beam deflection are outward from the tine's surface, resulting in a broadening of the tine in the image. Therefore, the observed inverse temporal breathing of the active and ground tines indicates that, upon the EM wave excitation an oscillating electric field perpendicular to $\vec{v}_e$ is built in the gaps between the active and ground tines. These images are a direct reflection of the EM wave propagation process through the interdigitated combs.

Considering a collimated beam illumination, the temporal electric field $E_{x,y}(x,y,t)$ around the tine is proportional to the change of the tine's edge-intensity profile in the time-frozen images (*27*), namely, the larger beam deflection means the larger local electric field. Shown in Fig. 3F are the time dependent imaging breathing (tine's edge variation) at three representative positions (P1, P2 and P3) around a ground tine, respectively, as indicated by the colored arrows in the top panel of Fig. 3D. All of them follows a same cosine function but with different amplitudes. Point 2 near the tine's corner exhibits a much bigger amplitude than other two positions, implying that there is a substantial local field enhancement at the corners of the tines in the EM waver propagation process, which will be discussed later.

We further studied the dependence of EM wave propagation dynamics on the excitation power. Additional ultrafast pump-probe imaging measurements were performed with different excitation powers from 0.5 W to 1.0 W, where the similar temporal breathing phenomenon of the active and ground tines was observed under different powers. Fig. 4A presents the plots of the time dependent width variation of a ground tine at all excitations, in which all the plots follow a sinusoidal function with the frequency of $5.25 \pm 0.02$ GHz (fitting in Fig. 4A) with no phase difference. While their amplitude increases with increasing the excitation power and follows a linear dependence (fit in Fig. 4B), i.e. the temporal oscillating electric field erected between the tines is linearly proportional to the excitation power within this power range.

To further understand the experimental observations, we performed numerical simulations on the EM wave propagation in a microstrip of two interdigitated combs with the same geometry and materials (Fig. S2). The simulation was carried out by a 3D electromagnetic finite element analysis package CST microwave studio (Materials and Methods). Hexahedron and local refine meshes were adopted to get high resolution EM field distribution along the sample. A frequency domain solver was used to solve the Maxwell's equation in the cells. The 5.25 GHz RF signal (power of 1.0 W) excites a traveling EM wave which propagates through the two interdigitated combs and is fully absorbed by the RF dump (load) at the end of the sample (Fig. S2).

Fig. 5A presents a set of typical snapshots of the simulated electric field distribution (projected in the *x-y* plane at the mid-comb thickness) around one active tine and two adjacent ground tines at different delay times (Movie S2), where the arrows indicate the direction of the fields and the field strength is encoded in the color. The sample is non-magnetic and the effects of magnetic fields are negligible in the experiment, which are not considered here. Clearly, as the EM wave propagates through the interdigitated combs under investigation, a temporal oscillating electric field $E_{x,y}(x,y,t)$ is instantly established in between the gaps (in the *x*-*y* plane) of the active and ground tines, and the electric field is perpendicular to the tine's surface along the beam direction. Specifically, with time elapses from 0 to 95 ps, the fields point from the active tine towards the neighboring ground ones, and gradually grow from zero to a maximum value at ~48 ps ($|E_x| \approx 1.7$ V/m) and then return back to zero at ~95ps. Further from 95 to 190s, the electric fields switch the direction and gradually increase to a maximum value at ~143 ps ($|E_x| \approx 1.7$ V/m) and then declines to zero again at ~190 ps. This process is repeated with each EM wave cycle. This temporal oscillating electric fields would exert a local field-dependent momentum kick on the imaging electrons that is proportional to the local waveform, resulting in the beam deflection and the breathing of the tines in the time-frozen images observed in the experiment. To show more clearly the temporal evolution of the field distribution, we plot the electric fields $E_x$ and $E_y$ as a function of time at three positions near a ground tine (P1, P2 and P3, in line with that in the experiment shown in Fig.3D) in Fig. 5B and 5C, respectively. The electric fields at all the three positions oscillating in a sinusoidal function with the frequency of 5.25 GHz, but different in the field strength amplitudes. In particular, the field strength ($E_x$) near the tine's corner is stronger than other two positions. At the position of P1, $E_y$ is nearly zero; while at the position of P3, $E_y$ is almost zero. These results demonstrate that the electric field vectors are vertically polarized to the surface of the tines and undergo a synchronously oscillation both in direction and strength with time, while the corners of each tine exhibit a substantial local field enhancement. To see more clearly the local field enhancement, we plotted the electric field strength $|E_x|$ (in absolute value) at *t* = 20 ps as a function of position near the surface of a ground tine in Fig. 5D (2D map of the field strength is shown in Fig. S3), where the positions are indicated by the red line with an arrow in the inset. As the position moves along the pink arrow, the field strength shows no apparent change in the parallel gap (from 0 to 20 μm) while exhibits a sharp increase near the corner (position P2 indicated by the pink arrow) and then gradually decreases to zero, indicating the existence of a remarkable non-linear local field enhancement at the singular points of the microstrip in the EM wave propagation process. This non-linear local field enhancement is caused by the convex surface geometry, where the smaller radius of the curved surface will result in a higher density of the equipotential surfaces and thus a larger local electric field. The results of the simulation are in good agreement with the experimental observations.

In summary, we developed a laser-free UEM with high spatiotemporal resolutions by integrating a RF-driven pulser to a commercial TEM, which allows facile operation in both the UEM mode and the conventional TEM mode. It offers a universal methodology for EM wave excitation and structure dynamic studies in real time and space by employing a straightforward retrofit. We used the laser-free UEM to study the GHz EM wave propagation dynamical process in a microstrip consisting of two interdigitated combs and demonstrated its ability for direct visualization of EM field oscillation with time, revealing field amplitude, polarization direction and wave propagation on the nanometer-ps time scale, which has not been accessible by other imaging methods. The demonstrated laser-free UEM provides a powerful methodology for real-space visualization of electrodynamics in small devices working with MHz to GHz operation frequencies, such as the collective carrier dynamics and field effects in miniaturized wireless antennas, sensors and RF MEMS (*46*). Future optimization of the input RF waveform and using a smaller chopping aperture could achieve sub-ps and even ~100 fs electron packets (*36, 37*), making fs time resolution for the laser-free UEM feasible. The laser-free UEM is also compatible with laser-triggered ultrafast pump-probe measurements using advanced laser-RF synchronization technologies (*44*). With these advanced features of the laser-free UEM, we envision the emergence of broad applications in many research areas, from materials physics to biology and mobile communication technologies.

## SUPPLEMENTARY MATERIALS

Supplementary materials are available in the online version of the paper, including Fig. S1 to S3, and Movie S1 to S2.

## Acknowledgements

This work was supported by the Materials Science and Engineering Divisions, Office of Basic Energy Sciences of the U.S. Department of Energy under Contract No. DESC0012704. The electron pulser was developed by Euclid through the DOE's SBIR grant under contract DE-SC0013121. X. F. is grateful to the financial support from the National Nature Science Foundation of China (NSFC) (No. 11974191).

## Author contributions

Y. Z., J. W. L. and X. F. conceived the research project. X. F., C. J., and Y. Z. did the experimental measurements. J. J. G. and V. J. prepared the sample. X. F. did the data analysis and wrote the manuscript with input from Y. Z., E. W. and C. J.. E. W. developed the model and performed the numerical simulations. All the authors contributed to the discussion and the revision of the manuscript.

## Competing financial interests

The authors declare no competing financial interests.

**Figure legends**

**Fig. 1. Laser-free UEM system.** (**A**) Schematic of the conceptual design of the laser-free UEM. Displayed is the TEM with the integration of a RF-driven pulser system and a frequency-double, delay-control RF circuit for the sample excitation. The pulser is inserted between the electron gun and the standard column lens. The inset shows a schematic design of the pulser, which consists of two traveling-wave metallic comb stripline elements: the modulator K1 and the demodulator K2, with a chopping aperture between them. The modulator K1 sweeps the continuous electron beam across the chopping aperture to create two electron pulses in each RF cycle, while the demodulator K2 compensates the K1-induced transverse momentum on the pulses to further rectify the shape of the chopped beam. (**B**) Photograph of our homebuilt laser-free UEM system based on a JEOL JEM-2100F Lorentz TEM. Displayed are the TEM with the RF-driven pulser inserted between the electron gun and the standard column lens and the connected RF source. The inset shows a picture of the modulator K1, the demodulator K2 and the chopping aperture inside the pulser.

**Fig. 2. Comparison of imaging and diffraction quality between the continuous beam mode and the pulsed beam mode.** Images and diffraction patterns acquired at the continuous beam mode: (**A**) Bright field image of gold nanoparticles; (**B**) Diffraction pattern of gold nanoparticles; (**C**) Diffraction pattern of a $VO_2$ single crystal (along [010] zone axis); (**D**) Out-of-focus Fresnel phase image of magnetic vortex in a circular ferromagnetic Permalloy disk. Images and diffraction patterns acquired at the pulsed beam mode with the repetition rate of 5.25 GHz: (**E**) Bright field image of gold nanoparticles; (**F**) Diffraction pattern of gold nanoparticles; (**G**) Diffraction pattern of a $VO_2$ single crystal (along [010] zone axis); (**H**) Out-of-focus Fresnel phase image of magnetic vortex in a circular ferromagnetic Permalloy disk.

**Fig. 3. Pump-probe imaging of the EM wave propagation dynamics in a microstrip of two interdigitated combs.** (**A**) Electron pulse duration as a function of the voltage amplitude $U_0$ of the input

RF source for the modulator K1. The red-dash-line is the curve fit with an inverse function of $\tau \propto 1/U_0$. (**B**) Schematic of the temporal oscillating electric field (normalized in the field strength) of a 5.25 GHz EM wave used for excitation. (**C**) Schematic of two interdigitated combs used for investigation (Fig. S1). The excitation signal is applied from one end of the two combs while the other end is terminated with a 50 Ω load to eliminate the EM wave reflections. (**D**) Typical snapshots of two ground tines and one active tine in between them at different delay times (Movie S1), obtained from the area indicated by the blue dashed box in Fig. 3C. The images are acquired at a magnification of 1200 X with an integral time of 1.5 s. The arrows indicate the initial positions of the two edges of each tine in the images without excitation. (**E**) The width variations (along $x$ direction) of the active (blue dots) and ground tines (red dots) due to the beam deflection as a function of delay time with curve fitting. They follow a cosine function and a sinusoidal function, respectively, with the same amplitude and frequency. (**F**) Time dependent imaging breathing of the tine's edge at three representative positions (P1, P2 and P3) around a ground tine, respectively, as indicated by the colored arrows in the first panel of Fig. 3D. Position P2 near the tine's corner exhibits a much higher breathing amplitude than other two positions, indicating a remarkable local field enhancement.

**Fig. 4. Excitation power dependence of the EM wave propagation dynamics.** (**A**) Plots of the time dependent width variation of a ground tine at different delay times with increasing excitation power from 0.5 W to 1.0 W. (**B**) The amplitude of the tine's width variation as a function of excitation power. It follows a linear power dependence.

**Fig. 5. Numerical simulations on the EM wave propagation dynamics in a two interdigitated combs**. (A) Typical snapshots of the simulated electric field distribution (projected in the x-y plane at the mid-comb thickness) around the active and ground tines at different delay times (Movie S2). The arrows indicate the direction of the electric fields with encoded color for the field strength. (**B**) Plots of the electric field $E_x$ as a function of time at three representative positions (P1, P2 and P3) around a ground tine. The field strength near the corner of the tine is stronger than other positions, indicating a local field enhancement near the corner. (C) Plots of the corresponding electric field $E_y$ as a function of time at the three representative positions. The field strength of $E_y$ at P1 is nearly zero and that of $E_x$ at P3 is almost zero, which indicates the established local field vectors are vertical to the tine's surfaces along the beam-pass direction. (**D**) Plot of the electric field strength of $|E_x|$ (in absolute value) as a function of position along the red line with an arrow (inset of Fig. 5D) near the surface of a ground tine. The sharp increase of the field strength near the corner (position P2) indicates a remarkable nonlinear local field enhancement. The field strength in the inset is color encoded with the color bar in the inset.

**Figures**

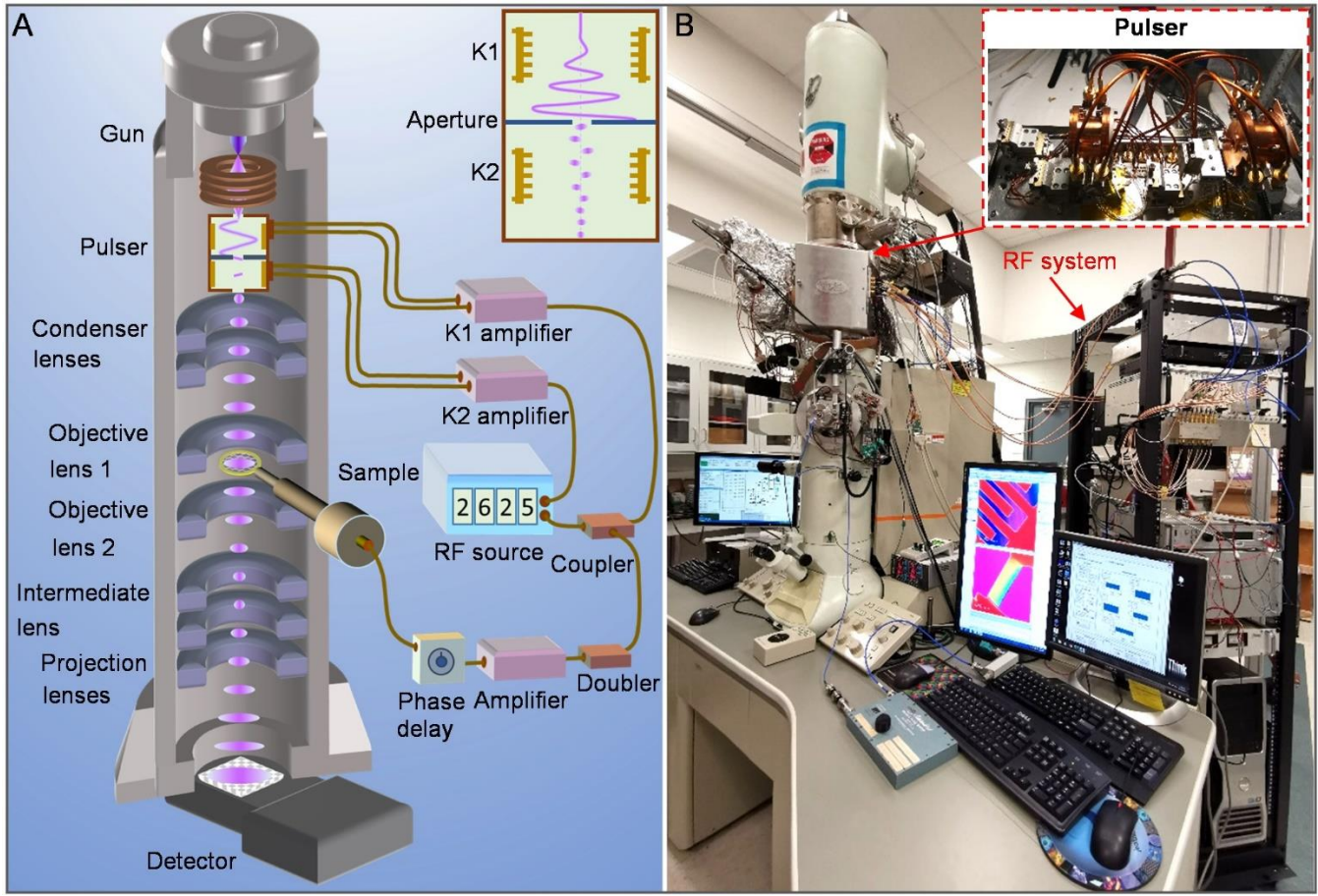

Figure 1

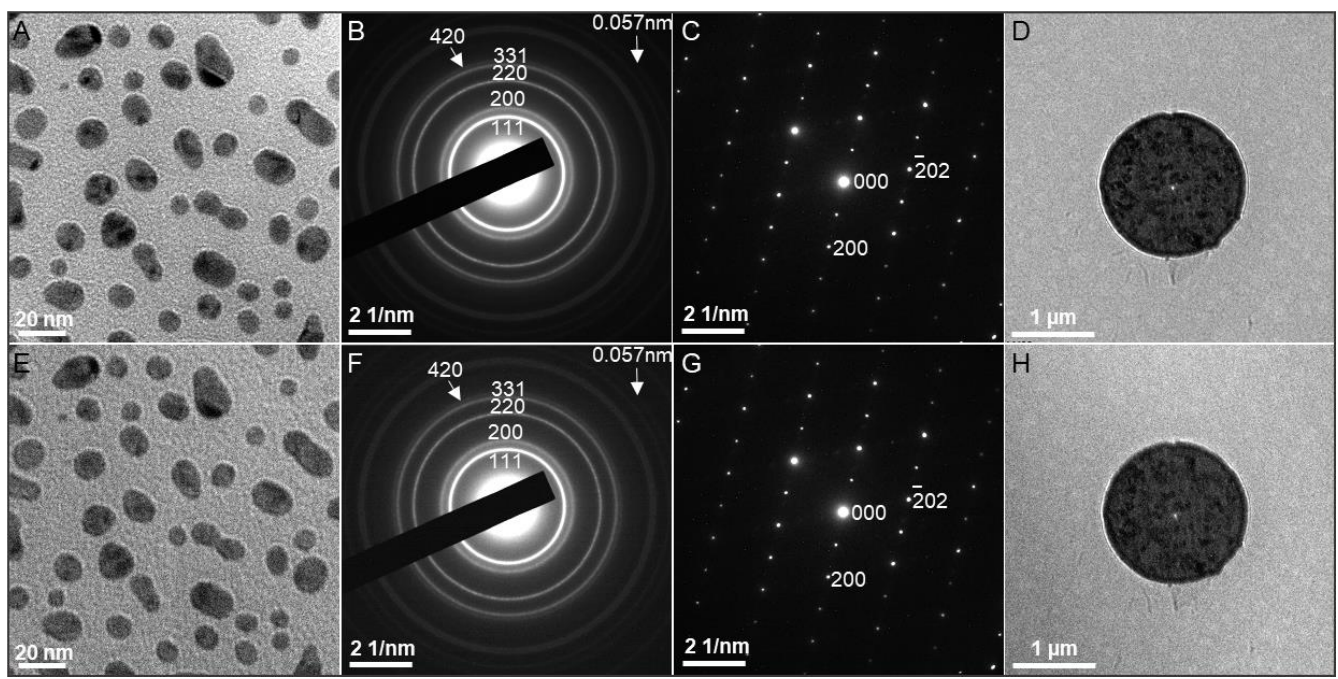

**Figure 2**

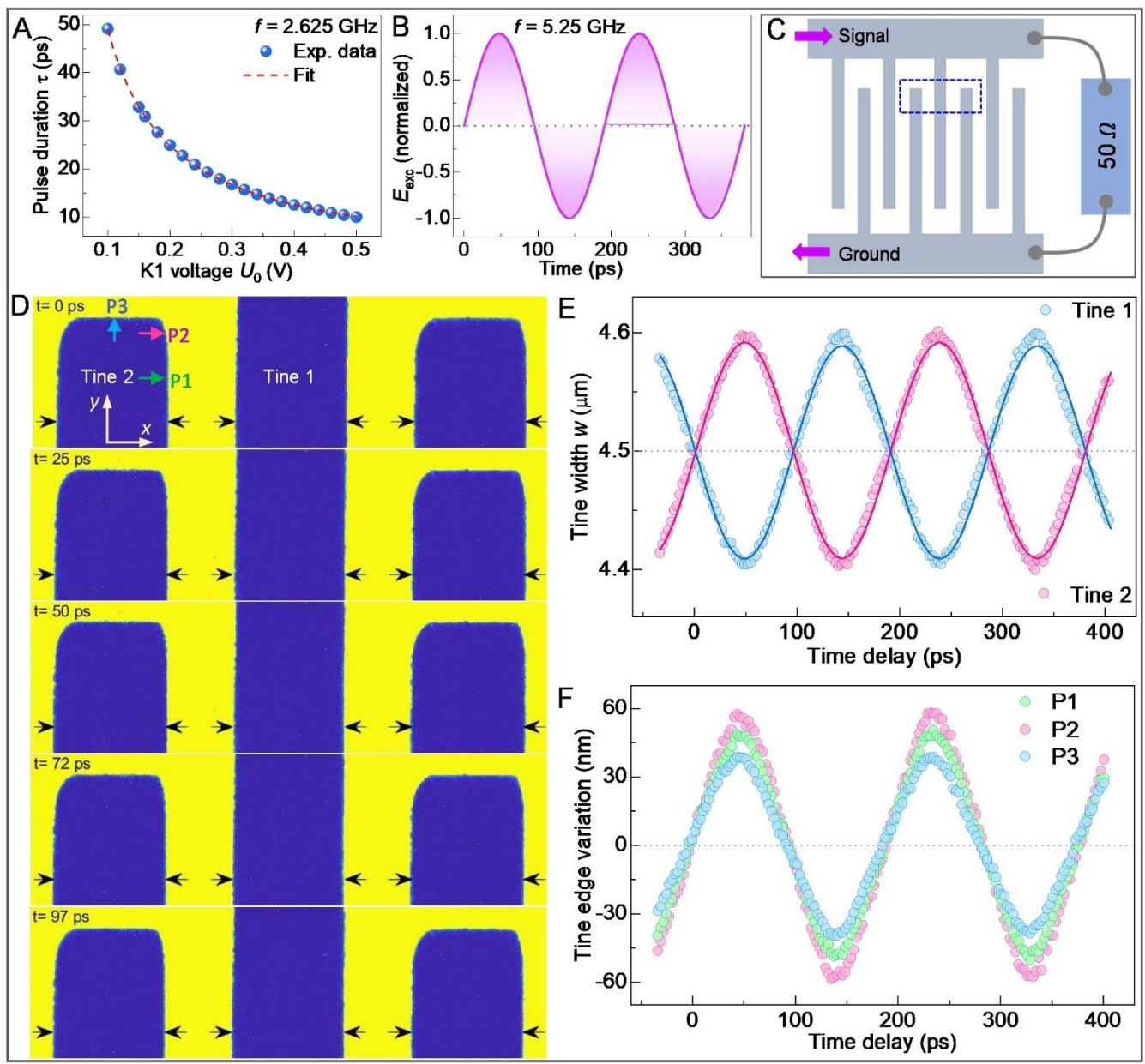

**Figure 3**

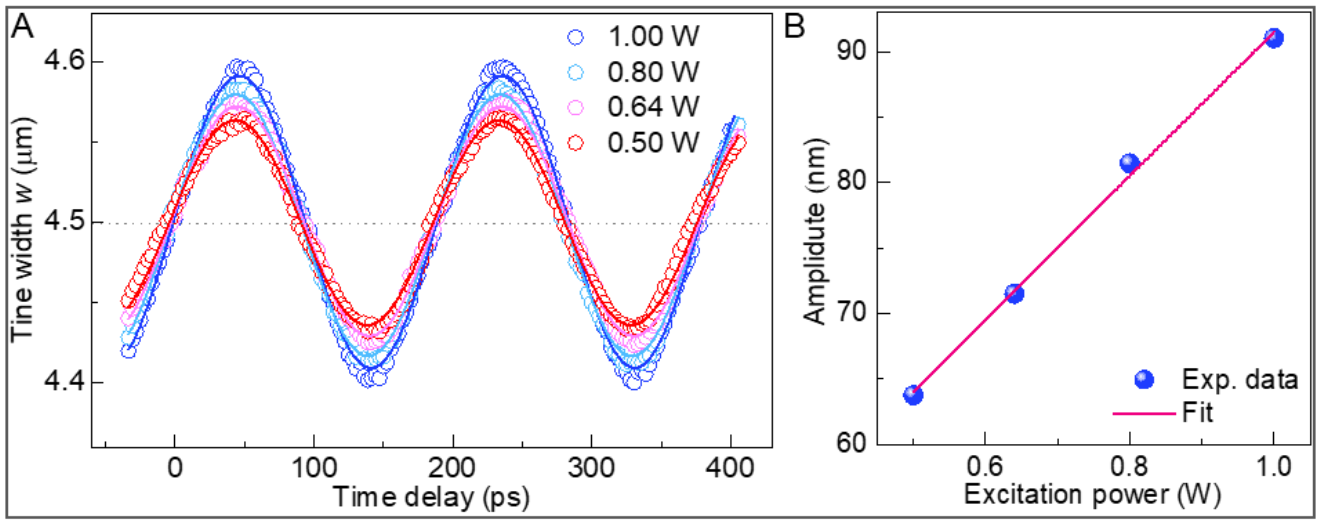

**Figure 4**

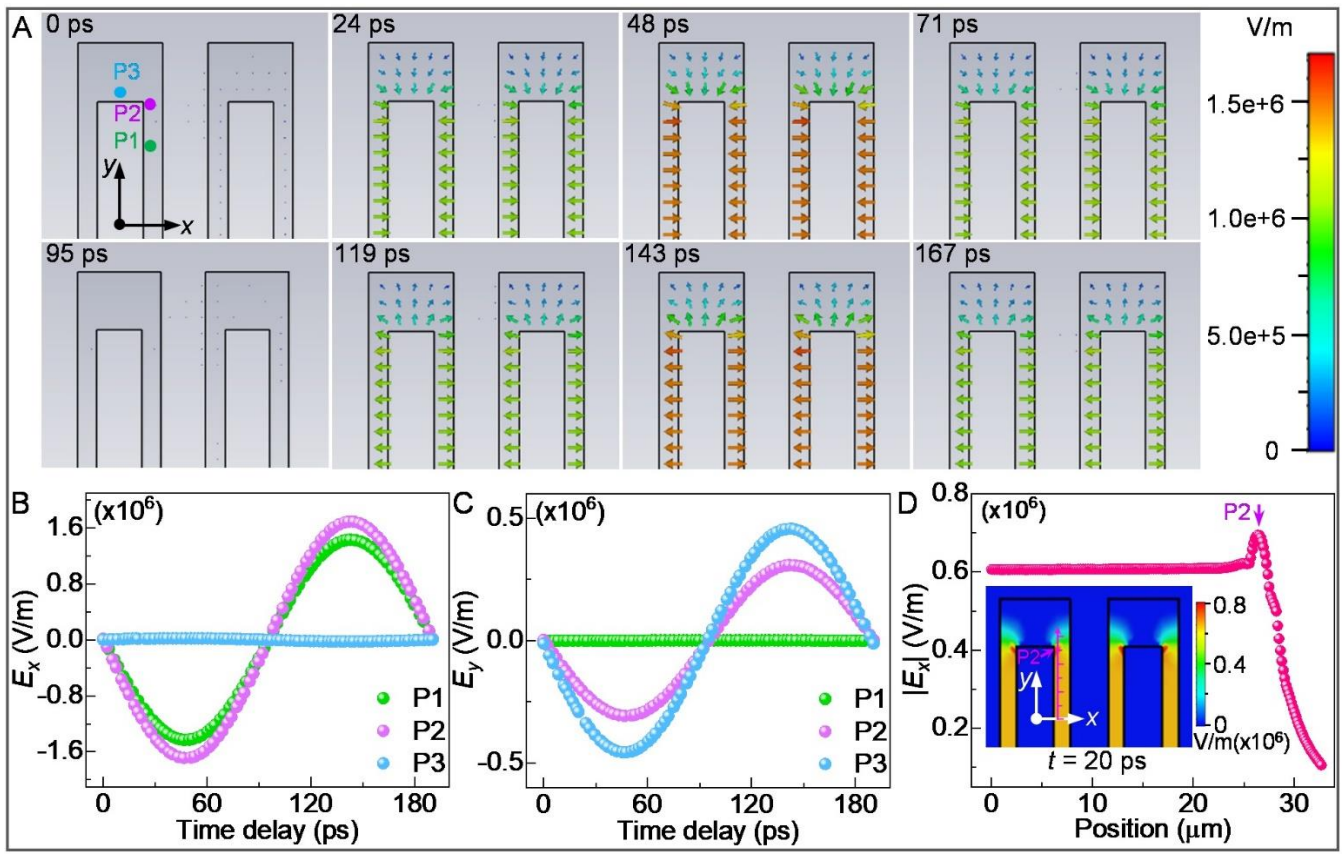

Figure 5